\documentclass[fleqn,10pt]{wlscirep}
\usepackage[utf8]{inputenc}
\usepackage[T1]{fontenc}
\usepackage{lineno}
\usepackage{multirow}
\usepackage{amsmath,amssymb}
\usepackage{fdsymbol}
\usepackage{makecell}
\usepackage{color,soul}

\title{Inflation of test accuracy due to data leakage in deep learning-based classification of OCT images}

\author[1,2,*]{Iulian Emil Tampu}
\author[1,2,3]{Anders Eklund}
\author[1,2]{Neda Haj-Hosseini}
\affil[1]{Department of Biomedical Engineering, Linköping University, 581 85 Linköping,  Sweden}
\affil[2]{Center for Medical Image Science and Visualization, Linköping University, 581 85 Linköping,  Sweden}
\affil[3]{Division of Statistics \& Machine Learning, Department of Computer and Information Science, Linköping University, 581 83 Linköping,  Sweden}

\affil[*]{corresponding author: Iulian Emil Tampu (iulian.emil.tampu@liu.se)}

\begin{abstract}
In the application of deep learning on optical coherence tomography (OCT) data, it is common to train classification networks using 2D images originating from volumetric data. Given the micrometer resolution of OCT systems, consecutive images are often very similar in both visible structures and noise. Thus, an inappropriate data split can result in overlap between the training and testing sets, with a large portion of the literature overlooking this aspect. In this study, the effect of improper dataset splitting on model evaluation is demonstrated for three classification tasks using  three OCT open-access datasets extensively used, Kermany’s and  Srinivasan's ophthalmology datasets, and AIIMS breast tissue dataset. Results show that the classification performance is inflated by 0.07 up to 0.43 in terms of Matthews Correlation Coefficient  (accuracy: 5\% to 30\%) for models tested on datasets with improper splitting, highlighting the considerable effect of dataset handling on model evaluation. This study intends to raise awareness on the importance of dataset splitting given the increased research interest in implementing deep learning on OCT data.

\end{abstract}
\begin{document}

\flushbottom
\maketitle

\thispagestyle{empty}

\section*{Introduction}
The evaluation of deep learning models, and machine learning methods in general, aims at providing an unbiased description of model performance. Given a pool of data suitable for studying a hypothesis (e.g., classification, regression or segmentation), different splits of the data are commonly created for model training, model hyper-parameter tuning (validation set) and model assessment (testing set). This translates to having a part of the data used to fit model parameters and tune model hyperparameters, and another set to assess model generalization on unseen data \cite{xu2018splitting}. Disregarding the choice of having separate validation and testing splits \cite{kuhn2013applied}, the strategy used to generate the testing set from the original pool of data can have a large impact on the assessment of the model's final performance. Several studies have investigated the effect of the relative size between training, validation and/or testing sets \cite{xu2018splitting, guyon1997scaling} on model performance, as well as how training and validation sets can be used via resampling techniques, such as cross-validation, during model training \cite{xu2018splitting, refaeilzadeh2009cross}. More importantly, it is widely accepted that no overlap should exist between the samples used for model fitting and hyperparameter tuning, and those belonging to the testing set. If overlap is present, the model performance will be biased and uninformative with respect to the generalization capabilities of the model to new samples. However, although trivial, when implementing data splitting strategies, the overlap between training and testing sets can be easily overlooked.  This is specifically more common in those applications where, due to hardware limitations or model design choices, the data from one subject (or acquisition) is used to obtain multiple dimensionally smaller samples used for model training and testing. An example of such a scenario is the slicing of a 3D volume into 2D images. In these cases, the overlap between training and testing sets results from having 2D images from the same subject (or acquisition) belonging to both sets. A proper splitting must therefore be performed on the volume or subject level and not on the slice level.

Nowadays, machine learning methods, especially convolutional neural networks (CNNs) and deep learning algorithms, are widely used in research to analyze medical image data. A plethora of publications describe CNN implementations on a variety of both 2D and 3D medical data \cite{litjens2017survey, ker2017deep, anwar2018medical}. Reliable evaluation of such methods is paramount since this informs the research community on the models’ performance, allows meaningful comparison between methods, and to a greater extent indicates which research questions might be worth further investigation. In this regard, many medical image analysis challenges were established that aim at providing an unbiased platform for the evaluation and ranking of different methods on a common and standard testing dataset. Despite the many limitations of the medical image analysis challenges, including missing information regarding how the ground truth was obtained, their contribution towards a more transparent and reliable evaluation of deep learning methods for medical image applications is valuable \cite{maier2018rankings}.

\begin{figure}[t]
\centering
\includegraphics[width=420pt]{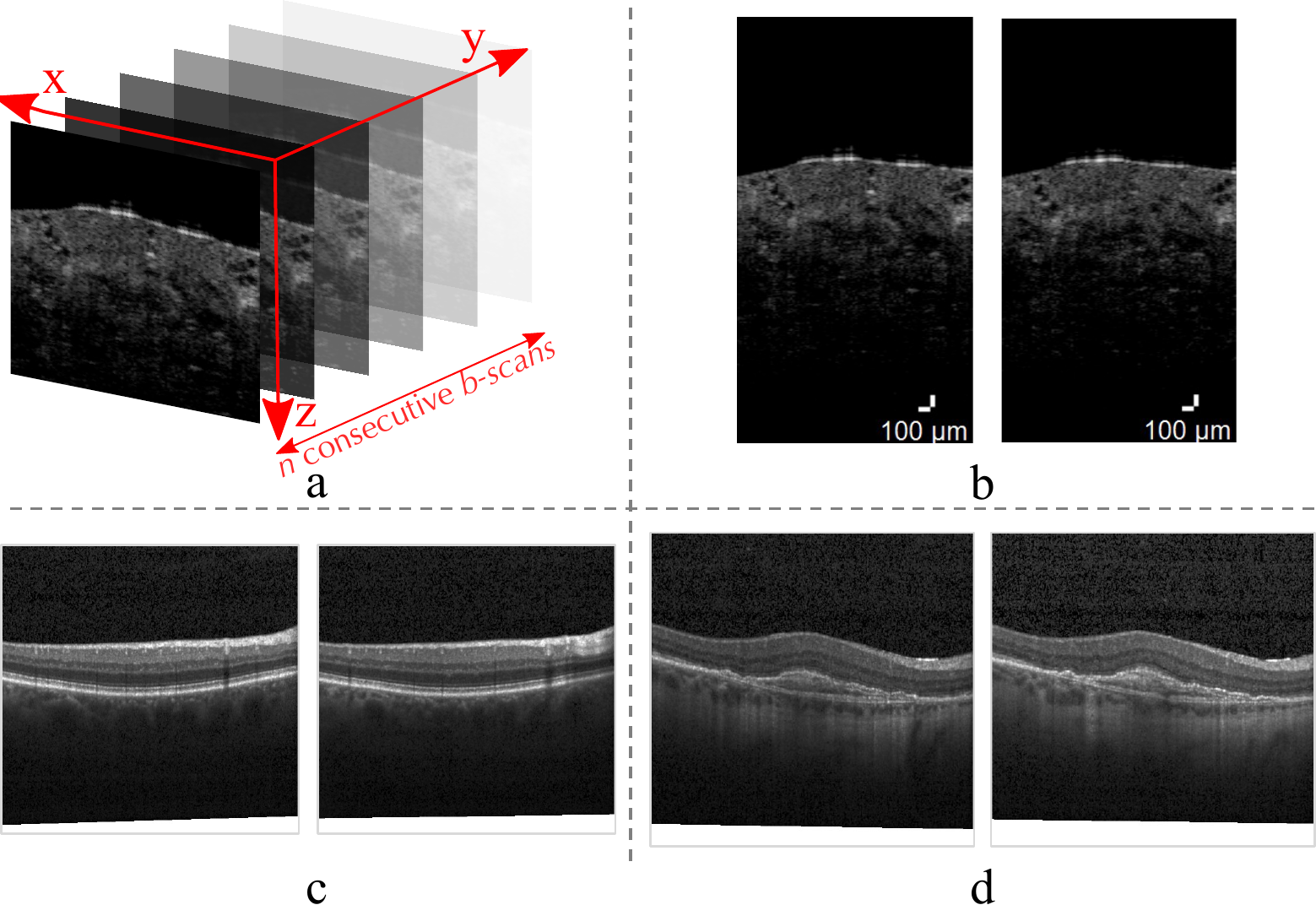}
\caption{Schematic of an OCT volume with examples of consecutive slices (\textit{b-scans}) from the three open-access OCT datasets used in this study. In (a) consecutive 2D \textit{b-scans} rendering a 3D OCT volume are pictured. Here an example from the AIIMS dataset \cite{butola2019volumetric} is used for illustrative purposes. In (b) the consecutive \textit{b-scans} separated by $\sim$18 micrometers are examples of healthy breast tissue (Patient 15, volume 0046, slices 0075 and 0076) from the AIIMS dataset \cite{butola2019volumetric}. In (c) images of healthy retina from Srinivasan's dataset \cite{srinivasan2014fully} (folder NORMAL5 slices 032 and 033) can be seen. In (d) consecutive images of retina affected by choroidal neovascularization (CNV 81630-33 and 81630-34) from the Kermany’s OCT2017 \textit{version 2} dataset \cite{kermany2018large} are shown. Note that the \textit{b-scans} from both Kermany’s and Srinivasan's datasets are given with data augmentation originally applied.}
\label{fig:introduction_1}
\end{figure}

However, not all implementations can be evaluated through dedicated challenges and unbiased datasets. Thus, when such third-party evaluation platforms are not available, it is the responsibility of the researchers performing the investigation to evaluate their \mbox{methods thoroughly}. As for the case of many of the reviewed medical image analysis challenges \cite{maier2018rankings}, one aspect that is sometimes missing or not well described is how the testing dataset is generated from the original pool of data. Moreover, there are examples where the preparation of the testing dataset was described, but its overlap with the training set was not considered \cite{wang2019deep, butola2020deep, irmak2021multi, sadad2021brain, yagis2021effect}, undermining the reliability of the reported results. Focusing on deep-learning applications for OCT, depending on the acquisition set-up, volumes are usually acquired with micrometer resolution in the x, y and z directions in a restricted field of view, with tissue structures that are alike and affected by similar noise. This results in consecutive slices having a high similarity, in both structure and noise. Figure \ref{fig:introduction_1} shows a schematic of an OCT volume along with examples of two consecutive slices from OCT volumes from three open-access datasets \cite{butola2019volumetric, kermany2018large, srinivasan2014fully}.

Most of the reviewed literature implementing deep learning on OCT data used 2D images from scanned volumes, where two methods were commonly used to split the pool of image data into training and testing sets: \textit{per-image} or \textit{per-volume/subject} splitting (Table \ref{tab:summary_literature}). 
In the \textit{per-volume/subject} splitting approach, the random selection of data for the testing set is done on the volumes (or subjects) ensuring that images from one volume (or subject) belong to either one of the training or testing sets. It is important to notice that even a \textit{per-volume} split might not be enough to avoid overlap between the training and testing set. In fact, if multiple volumes are acquired from the same tissue region, the tissue structures will be very similar among the volumes. In such scenarios, a \textit{per-subject} split is more appropriate. Overall, assuming that volumes are acquired from different tissue regions, splitting the dataset \textit{per-volume} or \textit{per-subject} (here called \textit{per-volume/subject}) ensures that overlap between training and testing is not present.
In the \textit{per-image} approach, 2D images belonging to the same volume are considered independent: thus, the testing set is created by random selection from the pool of images without considering that images from the same volume (or subject) might end up in both the training and testing sets. Even if this approach clearly results in an overlap between the testing and training sets, several of reviewed studies as well as an earlier version of one of the most downloaded open-access OCT datasets employed a \textit{per-image} split.

The aim of this study is to demonstrate the effect of improper dataset splitting (\textit{per-image}) on model evaluation using three open-access OCT datasets,  AIIMS \cite{butola2019volumetric}, Srinivasan's \cite{srinivasan2014fully} and Kermany’s OCT2017 (\textit{version 2} and \textit{3}) \cite{kermany2018large}. These were selected among the other open-access datasets for several reasons: (1) they are examples of OCT medical images belonging to different medical disciplines (ophthalmology and breast oncology) and showing different tissue structures and textures, (2) they are used in literature to evaluate deep learning-based classification of OCT images, with Kermany’s dataset \cite{kermany2018large} extensively used for developing deep learning methods in ophthalmology (over 14,500 downloads) \cite{Retinalkaggle}, and (3) the datasets are provided in two different ways, subject-wise for the AIIMS and Srinivasan's datasets, and already split for both versions of the Kermany’s datasets. The latter is an important aspect to consider since many of the studies using Kermany’s \textit{version 2} dataset overlooked the overlap between the training and the testing data. 

\begin{table}[t]
\centering
\caption{\label{tab:summary_literature} Summary of reviewed literature with a focus on dataset split and reported test classification performance. Open-access datasets and the ones available upon request are marked by $^\ast$ and $^{\ast\ast}$, respectively. The dataset is not open-access if not specified. Datasets obtained from animal model samples are marked by $^\dagger$. The difference in performance between studies using the same datasets results from the different methods implemented.}
\renewcommand{\arraystretch}{1.5}
\fontsize{9.8}{11}\selectfont 
\begin{tabular}{|c|l|l|l|}
\hline
\textit{Ref.} & \textit{OCT dataset} & \textit{Data split strategy} & \textit{Model performance on testing set} \\
\hline\hline
\cite{wang2019deep} & \makecell[l]{Thyroid, parathyroid, fat \\ and muscle samples} & \textit{per-image} &  97.12\% accuracy \\
\hline
\cite{micko2021diagnosis} & Pituitary adenoma & \textit{per-image} & 0.96 AUC\\
\hline
\cite{najeeb2018classification} & Ophthalmology \cite{kermany2018large}$^\ast$ (\textit{version 2}) & \textit{original split} & 95.55\% accuracy \\
\hline
\cite{chen2021classification} & Ophthalmology \cite{kermany2018large}$^\ast$ (\textit{version 2}) & \textit{original split} & 99.1\% accuracy \\
\hline
\cite{latha2021automated} & Ophthalmology \cite{kermany2018large}$^\ast$ (\textit{version 3}) & \textit{original split} & 98.7\% accuracy \\
\hline
\cite{kermany2018identifying} & Ophthalmology \cite{kermany2018large}$^\ast$ (\textit{version 3}) & \textit{original split} & 96.6\% accuracy \\
\hline
\cite{tsuji2020classification} & Ophthalmology \cite{kermany2018large}$^\ast$ (train \textit{version 2}, test \textit{version 3}) & \textit{original split}  & 99.6\% accuracy \\
\hline
\cite{kamran2019optic} & \makecell[l]{(1) Ophthalmology \cite{kermany2018large}$^\ast$ (train \textit{version 2}, test \textit{version 3}) \\(2) Ophthalmology \cite{srinivasan2014fully}$^\ast$} & \makecell[l]{(1) \textit{original split}\\(2) \textit{per-volume/subject}} & \makecell[l]{(1) 99.80\% accuracy\\(2) 100\% accuracy} \\
\hline
\cite{athanasiou2019deep} & Coronary artery & \textit{per-volume/subject} & 96.05\% accuracy \\
\hline
\cite{wang2021deep} & Kidney$^\dagger$ & \textit{per-volume/subject} & 82.6\% accuracy \\
\hline
\cite{gesperger2020improved} & High and low grade brain tumors & \textit{per-volume/subject} & 97\% accuracy \\
\hline
\cite{saratxaga2021characterization} & Colon$^{\ast \ast, \dagger}$ & \textit{per-volume/subject} & 88.95\% accuracy on 2D images \\
\hline
\cite{singla2019automated} & Breast tissue & \textit{per-volume/subject} & 91.7\% specificity \\
\hline
\cite{chetoui2020deep} & Ophthalmology \cite{kermany2018large}$^\ast$ (\textit{version 2}) & \textit{per-volume/subject} & 98.46\% accuracy\\
\hline
\cite{butola2020deep} & \makecell[l]{(1) Ophthalmology  \cite{kermany2018large}$^\ast$ (\textit{version 3}) \\(2) Ophthalmology \cite{srinivasan2014fully}$^\ast$ \\(3) Breast tissue \cite{butola2019volumetric}$^\ast$} & \makecell[l]{(1) \textit{original split} \\(2) \textit{per-volume/subject} \\(3) \textit{per-image}} & \makecell[l]{(1) $\sim$96\% accuracy \\ (2) > 98.8\% accuracy \\ (3) 98.8\% accuracy} \\
\hline
\cite{thomas2021novel} & \makecell[l]{(1) Ophthalmology \cite{srinivasan2014fully}$^\ast$ \\ (2) Ophthalmology \cite{rasti2017macular}$^\ast$ \\ (3) Ophthalmology  \cite{farsiu2014quantitative}$^\ast$ \\ (4) Ophthalmology  \cite{kermany2018large}$^\ast$ (unclear version)} & \makecell[l]{(1) \textit{per-volume/subject} \\ (2) \textit{per-volume/subject} \\ (3) \textit{per-volume/subject} \\ (4) \textit{original split}} & \makecell[l]{(1) 96.66\% accuracy \\ (2) 98.97\% accuracy \\ (3) 99.74\% accuracy \\ (4) 99.78\% accuracy} \\
\hline
\cite{karimian2018deep} & Dentistry & No description given & \makecell[l]{98\% sensitivity \\ 100\% specificity} \\
\hline
\cite{wang2020oct} & Ophthalmology & No description given & 99.19\% accuracy\\
\hline
\end{tabular}
\end{table}

\section*{Results}
\begin{figure}[t]
\centering
\includegraphics[width=410pt]{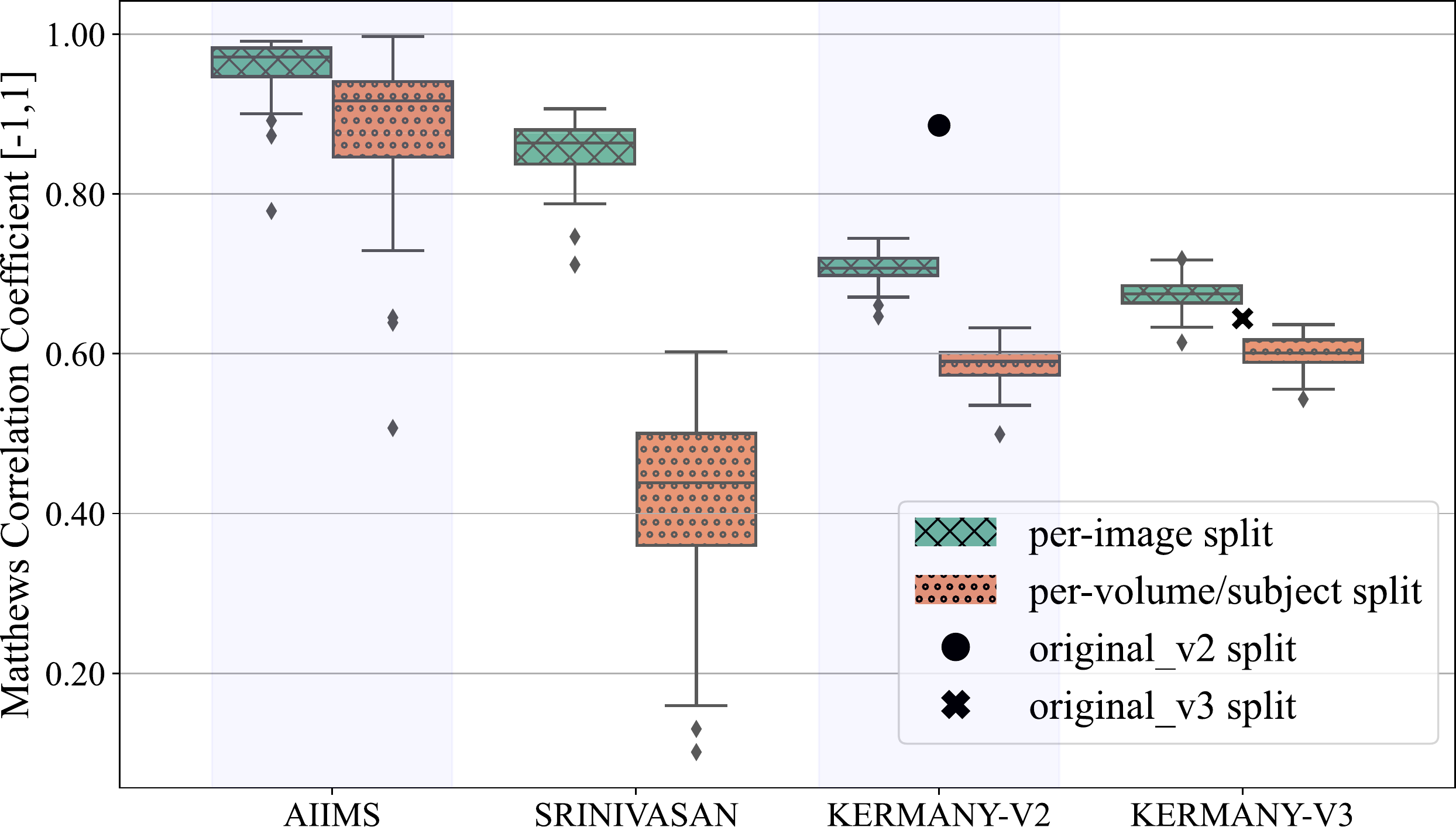}
\caption{Comparison between Matthews Correlation Coefficient for LightOCT model trained on different dataset split strategies. Each box plot summarizes the test MCC for the 50 models trained through a ten-times repeated five-fold cross validation.  Results are presented for all the four datasets with the \textit{per-image} split strategy shown in striped-green and \textit{per-volume/subject} split strategy in dotted-orange. For Kermany's datasets, the result of the models trained on the \textit{original\_v2} and \textit{original\_v3} splits are shown as full-black circle and full-black cross, respectively. Outliers are shown as diamond ($\blacklozenge$).}
\label{fig:results_1}
\end{figure}

LightOCT model classification performance on the three datasets and for the different dataset split strategies is summarized in Table \ref{tab:results_1}, with results presented as mean$\pm$standard deviation (m$\pm$std) over the ten-times repeated five-fold cross validation. In addition, Figure \ref{fig:results_1} shows the Matthews Correlation Coefficient (MCC) distribution as box plots.  For all datasets, the \textit{per-image} split strategy results in a higher model performance compared to the \textit{per-volume/subject} strategy.  Looking at both the AIIMS and Srinivasan's datasets, a model trained using a \textit{per-image} strategy had a higher mean MCC by 0.08 and 0.43, respectively,  compared to the one trained on a \textit{per-volume/subject} split.  A similar trend can be seen for both versions of Kermany's dataset, with an increase in mean MCC by 0.12 and 0.07 for \textit{version 2} and \textit{version 3}, respectively, when shifting from a \textit{per-volume/subject} to a  \textit{per-image} split strategy. Moreover, results on the original splits for both versions of Kermany's dataset are higher compared to the corresponding \textit{per-volume/subject} splits, with a difference in MCC of 0.30 and 0.04 for the \textit{version 2} and \textit{version 3} datasets, respectively.

Results on the random label experiments show that, for the original Kermany \textit{version 2} dataset, the obtained \textit{p}-value was 0.071, with the high MCC for random labels indicating a probable data leakage. For the other datasets, the \textit{p}-values were much larger. We conclude that using random labels during training can potentially be a way to automatically detect data leakage, but that it requires further research.

\begin{table}[t]
\centering
\caption{\label{tab:results_1}LightOCT model performance on the AIIMS \cite{butola2019volumetric}, Srinivasan's \cite{srinivasan2014fully} and Kermany’s \cite{kermany2018large} datasets with training, validation and testing sets split using different strategies. Performance metrics are reported as mean$\pm$standard deviation (m$\pm$std) over the models trained through ten-times repeated five-fold cross validation and classes, for the \textit{per-image} and \textit{per-volume/subject} splits.  For the original splits given by Kermany, results are reported for the single given split. AUC: area under the receiver operating characteristic curve, MCC: Matthews Correlation Coefficient.}
\fontsize{7.8}{11}\selectfont 
\renewcommand{\arraystretch}{1.5}
\begin{tabular}{|c|c|c|c|c|c|c|c|}
\hline
\textit{Dataset} & \textit{Split strategy} &  \makecell{ \textit{MCC} [-1,1] \\ \textit{(m$\pm$std)}} &  \makecell{ \textit{AUC} [0,1] \\ \textit{(m$\pm$std)}} &  \makecell{\textit{F1-score} [0,1] \\ \textit{(m$\pm$std)}} &  \makecell{\textit{Accuracy} [0,1] \\ \textit{(m$\pm$std)}} &  \makecell{\textit{Precision} [0,1] \\ \textit{(m$\pm$std)}} & \makecell{\textit{Recall} [0,1]\\ \textit{(m$\pm$std)}} \\
\hline\hline
\multirow{2}{*}{\makecell{AIIMS \cite{butola2019volumetric}} } & \textit{per-image} & \small{0.958﻿$\pm$0.038﻿}  & \small{1.000﻿$\pm$0.000﻿}  & \small{0.978﻿$\pm$0.021﻿}  & \small{0.978﻿$\pm$0.021﻿}  & \small{1.000﻿$\pm$0.000﻿}  & \small{0.978﻿$\pm$0.021﻿﻿} \\
\cline{2-8}
& \textit{per-volume/subject} & \small{0.881﻿$\pm$0.102﻿}  & \small{0.996﻿$\pm$0.009﻿}  & \small{0.934﻿$\pm$0.063﻿}  & \small{0.935﻿$\pm$0.060﻿}  & \small{0.993﻿$\pm$0.014﻿}  & \small{0.935﻿$\pm$0.060﻿﻿} \\
\hline\hline
\multirow{2}{*}{\makecell{Srinivasan \cite{srinivasan2014fully}} }  & \textit{per-image} &  \small{0.853﻿$\pm$0.039﻿}  & \small{0.985﻿$\pm$0.005﻿}  & \small{0.898﻿$\pm$0.030﻿}  & \small{0.899﻿$\pm$0.029﻿}  & \small{0.973﻿$\pm$0.009﻿}  & \small{0.899﻿$\pm$0.029﻿﻿} \\
\cline{2-8}
&  \textit{per-volume/subject} & \small{0.426﻿$\pm$0.116﻿}  & \small{0.817﻿$\pm$0.055﻿}  & \small{0.593﻿$\pm$0.088﻿}  & \small{0.603﻿$\pm$0.078﻿}  & \small{0.702﻿$\pm$0.078﻿}  & \small{0.603﻿$\pm$0.078﻿﻿}  \\
\hline\hline
\multirow{3}{*}{\makecell{Kermany \cite{kermany2018large} \\ \textbf{\textit{version 2}}} } & \textit{original\_v2} & \small{0.886}  & \small{0.993﻿}  & \small{0.909﻿}  & \small{0.911﻿} & \small{0.983﻿} & \small{0.911﻿} \\
\cline{2-8}
& \textit{per-image} & \small{0.707﻿$\pm$0.021﻿}  & \small{0.953﻿$\pm$0.003﻿}  & \small{0.764﻿$\pm$0.022﻿}  & \small{0.770﻿$\pm$0.019﻿}  & \small{0.886﻿$\pm$0.007﻿}  & \small{0.770﻿$\pm$0.019﻿﻿} \\
\cline{2-8}
& \textit{per-volume/subject} \ & \small{0.588﻿$\pm$0.025﻿}  & \small{0.890﻿$\pm$0.006﻿}  & \small{0.644﻿$\pm$0.033﻿}  & \small{0.669﻿$\pm$0.023﻿}  & \small{0.769﻿$\pm$0.012﻿}  & \small{0.669﻿$\pm$0.023﻿﻿} \\
\hline\hline
\multirow{3}{*}{\makecell{Kermany \cite{kermany2018large} \\ \textbf{\textit{version 3}}} } & \textit{original\_v3} & \small{0.644﻿}  & \small{0.964﻿}  & \small{0.678﻿}  & \small{0.704﻿} & \small{0.916﻿} & \small{0.704﻿} \\
\cline{2-8}
& \textit{per-image} &  \small{0.673﻿$\pm$0.021﻿}  & \small{0.950﻿$\pm$0.003﻿}  & \small{0.729﻿$\pm$0.022﻿}  & \small{0.738﻿$\pm$0.019﻿}  & \small{0.886﻿$\pm$0.007﻿}  & \small{0.738﻿$\pm$0.019﻿﻿}  \\
\cline{2-8}
& \textit{per-volume/subject} &  \small{0.600﻿$\pm$0.021﻿}  & \small{0.911﻿$\pm$0.006﻿}  & \small{0.651﻿$\pm$0.028﻿}  & \small{0.671﻿$\pm$0.021﻿}  & \small{0.795﻿$\pm$0.012﻿}  & \small{0.671﻿$\pm$0.021﻿﻿} \\
\hline
\end{tabular}
\end{table}

\section*{Discussion}
Dataset split should be carefully designed to avoid overlap between training and testing sets. Table \ref{tab:summary_literature} summarizes several studies on deep learning applications for OCT data, specifying the described data split strategy.  All the works using a \textit{per-image} split strategy or the original split from both versions of Kermany's dataset reported accuracies >95\%. The results obtained on Kermany's \textit{original\_v2} split are in accordance with reported accuracy \cite{najeeb2018classification, chen2021classification, chetoui2020deep},  where the differences in performance can be attributed to the deep learning model used and its optimization. However, results on the third version of the dataset were much lower compared to those in literature \cite{butola2020deep,latha2021automated,kermany2018identifying}, even when using the same model architecture, loss and optimizer \cite{butola2020deep}. Interestingly, Table \ref{tab:summary_literature} also shows that studies using multiple datasets and reporting different split strategies for each dataset \cite{butola2020deep,kamran2019optic, thomas2021novel}, report as high accuracy on the \mbox{\textit{per-volume/subject}} split datasets as the one on the \textit{original\_v2} or the \textit{per-image} split datasets. Moreover, the drop in MCC seen when using Kermany's \textit{original\_v2} compared to the \textit{original\_v3} split, highlights the inflation effect that leakage between training and testing data has on model performance, especially when the \textit{original\_v2} shows to have 92\% overlap between training and testing sets while \textit{original\_v3} has none. In the light of our results, it is reasonable to question whether the high performances reported for the \textit{per-volume/subject} split datasets reflect the true high performance of the implemented methods, or are examples of inflated accuracy values due to data leakage between training and testing sets.

The difference in the effect of data leakage between datasets can be attributed to two reasons: (1) the way the dataset is provided for download and (2) the classification task that the model needs to learn.  Firstly, Kermany's dataset is provided already split in training, testing and validation sets (\textit{version 2} of the dataset), with overlap at both subject-ID level and at image level. Thus, using the \textit{original\_v2} dataset as provided, highly inflates the model performance, with a drop in performance seen when using a \textit{per-image} split (overlap only on a subject-ID level not on an image level) and an even larger decrease in the case of the appropriate split. AIIMS and Srinivasan's datasets are provided with data from each subject saved independently, with split being performed in the reviewed studies using a \textit{per-image} or \textit{per-subject/volume} strategy (see Table \ref{tab:summary_literature} ). Thus, for these datasets the effect of inflation was measured with respect to the overlap in subject-IDs between the training and testing sets, and not with respect to having the exact same image in both sets. Disregarding the way the datasets are provided, the second reason for which the effect of data leakage was different between the datasets could be attributed to the difficulty of the classification task. Looking at the results for the proper split, the model performance on AIIMS dataset was higher compared to the one for Kermany's dataset, suggesting that the binary classification between cancer and normal breast tissue was easier compared to the four-class classification for Kermany's dataset. When improperly splitting the data, the advantage given to the model in classifying the test data by training on samples in the testing set had a smaller impact on AIIMS dataset than for Kermany's since the classification task was easier. However, it is not trivial what task can be considered easy since this is influenced by a multitude of factors, including data, model architecture, and model optimization.

This study is limited in investigating only data leakage originating from the overlap between the training and testing sets. However, there are other sources of data leakage stemming from hyper-parameter optimization, data normalization, and data augmentation. During hyper-parameter optimization, model parameters as well as training strategies should not be tuned based on test performance. Similarly, in the case of data normalization, statistics such as mean and variance used for the normalization should only be computed  on the training set, and not the entire dataset.  Failing to do so results in biasing method design choices with information obtained from the testing set, thus compromising the evaluation of model generalization to new data. Finally, in leakage due to data augmentation, an image could be augmented multiple times, with its different augmented versions ending up in the training and testing set. This results in a data overlap similar to the one of a \textit{per-image} split strategy, where images with the same structures and noise properties are in both training and testing sets. Both original splits in the Kermany and Srinivasan's datasets are provided with images already augmented; however, overlap between training and testing with respect to data augmentation was not investigated in this work.

In conclusion, the dataset split strategy that is used can have a substantial impact on the evaluation of deep learning models. In this paper, it is demonstrated that, in OCT image classification applications specifically, a \textit{per-image} split strategy of the volumetric data adopted by a considerable number of studies, returns over-optimistic results on model performance and an inflation of test performance values. This calls into question the reliability of the assessments and hindering an objective comparison between research outcomes. This problem has also been demonstrated in 3D magnetic resonance (MR) imaging studies \cite{yagis2021effect} and in digital pathology \cite{bussola2021ai}, where data leakage between the training and testing sets resulted in over-optimistic classification accuracy (>29\% slide level classification accuracy in MR studies and up to 41\% higher accuracy in digital pathology). Moreover, greater attention should be paid to the structure of datasets made available to the research community to avoid biasing the evaluation of different methods and undermining the usefulness of open-access datasets. With the increased interest of the research community in the use of deep learning methods for OCT image analysis, this study intends to raise awareness on a trivial but overlooked problem that can spoil research efforts if not addressed correctly.

\section*{Methods}
\subsection*{Dataset description}

\subsubsection*{AIIMS dataset}
The AIIMS dataset is a collection of 18480 2D OCT images of healthy (n=9450) and cancerous (n=9030) breast tissue \cite{butola2019volumetric}. The images are obtained from volumetric acquisitions and are provided as BMP files of size 245$\times$442 pixels organized per-class and per-subject (22 cancer subjects and 23 healthy subjects). 
\subsubsection*{Srinivasan's dataset}
The Srinivasan's ophthalmology dataset \cite{srinivasan2014fully} collects a total of 3,231 2D OCT images of age-related macular degeneration (AMD), diabetic macular edema (DME), and normal subjects.  For each class, data from 15 subjects is provided in independent folders. OCT images are given as TIFF files with 512$\times$496 pixels saved after data augmentation (rotation and horizontal flip).
\subsubsection*{Kermany's dataset}
The Kermany's ophthalmology dataset \cite{kermany2018large, kermany2018identifying} is one of the largest open-access ophthalmology datasets\cite{khan2021global} and is used by an extensive number of studies (see Table \ref{tab:summary_literature}). The dataset contains images from 5319 patients (\textit{train}=4686, \textit{test}=633) for retina affected by choroidal neovascularization (CNV), diabetic macular edema (DME), and drusen as well as from normal retina. The dataset is available in different versions with \textit{version 2} and \textit{version 3} (latest) both used in literature.  The difference between the dataset versions is threefold: (1) the total number of available images and their organization, (2) the number of images in the given testing set and (3) the extent of data overlap between the given training and testing sets. For both dataset versions, images are given as JPEG files of sizes ranging [384 to 1536]$\times$[496 to 512] pixels saved after data augmentation (rotation and horizontal flip). The \textit{version 2} of the dataset is provided with splits for training (\textit{n}=83484 images), validation (\textit{n}=32 images), and testing sets (\textit{n}=968 images),  with validation and testing sets balanced with respect to the classes.  The \textit{version 3} of the dataset is given with training (\textit{n}=108312) and testing (\textit{n}=1000, balanced between the four classes).  In this study, the given splits are referred to as \textit{original\_v2} and \textit{original\_v3} splits, for \textit{version 2} and \textit{version 3} of the dataset, respectively.
For both versions of the dataset, there is no specification on if the split between sets is performed before or after data augmentation as well as if the split in training and testing sets was performed \textit{per-image} or \textit{per-volume/subject}.  By performing an automatic check on the \textit{original\_v2} split (assuming that the naming convention is CLASS\_subject-ID\_bscan-ID), it was found that 92\% of the test images belong to subject-IDs also found in the training set. Moreover, by visually inspecting the given splits it was possible to identify images in the testing set that were similar to the training set (an example of such a case is training image=DRUSEN-8086850-6, testing image=DRUSEN-8086850-1). When performing the same automatic check on the \textit{original\_v3} split, no overlap between subject-IDs was found.  In Table \ref{tab:summary_literature} it is specified which version of Kermany's dataset was used by the different studies. Among these, two studies used a mixture of both datasets \cite{tsuji2020classification, kamran2019optic}.  

\subsubsection*{Dataset split}
For all datasets, a custom split function was implemented to split the dataset \textit{per-image} or \textit{per-volume}. In either case,  1000 images from every class were assigned for testing in the case of Kermany's and the AIIMS datasets. For the Srinivasan's dataset, 250 images were selected for testing instead, given the smaller number of total images.  Example images from AIIMS, Srinivasan's and Kermany's \textit{version 2} datasets are shown in Figures \ref{fig:introduction_1}b,  \ref{fig:introduction_1}c and \ref{fig:introduction_1}d, respectively.

\subsection*{Model architecture and training strategy}
The LightOCT model proposed by Butola et al. \cite{butola2020deep} was used in this study. LightOCT is a custom, shallow and multi-purpose network for OCT image classification composed of a two-layer CNN encoder, and one fully connected layer with softmax activation as output layer. The first and second convolutional layers have 8 and 32 convolutional filters, respectively. The kernel size of the filters in both layers is set to 5$\times$5 and the output of each layer passes through a ReLU activation function \cite{butola2020deep}. A max-pooling operation is present between the first and the second convolutional layer that cuts the spatial dimension of the output of the first layer in half. The two-dimensional output of the CNN encoder is then flattened to a one dimensional vector, which is fed to the fully connected layer for classification. The number of nodes in the fully connected layer is changed based on the number of classes specified by the classification task \cite{butola2020deep}. 

For all of the classification tasks, the model was trained from scratch using stochastic gradient descent with momentum (m=0.9) with a constant learning rate (lr=0.0001). For all experiments, the batch size was set to 64 and the model was trained for 250 epochs without early stopping. Note that model architecture and training hyperparameters were not optimized for each dataset since it was out of the scope of this work. The model architecture as well as the training hyperparameters were chosen based on the results of Butola et al.\cite{butola2020deep} The model and the training routine were implemented in Tensorflow 2.6.2, and training was run on a computer with a 20-core CPU and 4 NVIDIA Tesla V100 GPUs. 

\subsection*{Evaluation metrics}
Models were trained on the original splits, if available, and on training and testing splits obtained using a \textit{per-image} and \textit{per-volume/subject} strategy. A ten-times repeated five-fold cross validation was run for both split strategies to ensure reliability of the presented results \cite{maqc2010microarray}.  A multi-class confusion matrix was used to evaluate the classification performance of the model with Matthews Correlation Coefficient (MCC) obtained as a derived metric coherent with respect to class imbalance and stable to label randomization \cite{jurman2012comparison, chicco2020advantages,chicco2021matthews}.  Accuracy, precision, recall and F1-score were also derived for each class using the definitions provided by Sokolova et al. \cite{sokolova2009systematic} to allow comparison with previous studies.  Additionally, receiver operator characteristic (ROC) curves were used along with the respective area under the curve (AUC).
In an attempt to automatically detect bias due to data overlap, a random label experiment was carried out, where random labels were used for training a classifier and MCC was calculated on the test set (with original labels). To determine if the obtained MCC value was within the expected range, a null distribution was created for each dataset by creating 10000 test labels and prediction sets (also in this case random labels were used to simulate the random distribution) and calculating MCC for each of them. The MCC values obtained from the models trained using random labels on the different data splits (for which some have overlap) were compared to the respective null distribution (built without overlap) to calculate p-values using the one-sample Wilcoxon test (two-tailed).

\section*{Acknowledgments}
The study was supported by the grants from Åke Wiberg Stiftelse (M19-0455, M20-0034, M21-0083), FORSS - 931466, Vinnova project 2017-02447 via Medtech4Health and Analytic Imaging Diagnostics Arena (1908), Swedish research council (2018-05250) and ITEA / VINNOVA funded project ASSIST (2021-01954).

\section*{Dataset availability statement}
The datasets used in this study are open-access, with the AIIMS dataset \cite{butola2019volumetric} available at \url{https://www.bioailab.org/datasets}, Srinivasan's at \url{https://people.duke.edu/~sf59/Srinivasan_BOE_2014_dataset.htm}, Kermany’s  OCT2017 \cite{kermany2018large}  \textit{version 2} at \url{https://data.mendeley.com/datasets/rscbjbr9sj/2} and \textit{version 3} at \url{https://data.mendeley.com/datasets/rscbjbr9sj/3}.

\section*{Code availability}
The code used to generate the results in this paper is available at \url{https://github.com/IulianEmilTampu/SPLIT_PROPERLY_OCT_DATA.git}

\section*{Author contributions statement}
IET contributed with conceptualization, methodology, code development and implementation, formal analysis and drafting the manuscript. AE contributed with supervision and hardware resources. NHH contributed with conceptualization, supervision and funding. All authors contributed to the interpretation of the results, have revised and edited the manuscript and approved the submitted version.

\section*{Competing interests}
AE has previously received NVIDIA hardware for research.

\bibliography{references.bib}

\begin{thebibliography}{10}
\urlstyle{rm}
\expandafter\ifx\csname url\endcsname\relax
  \def\url#1{\texttt{#1}}\fi
\expandafter\ifx\csname urlprefix\endcsname\relax\def\urlprefix{URL }\fi
\expandafter\ifx\csname doiprefix\endcsname\relax\def\doiprefix{DOI: }\fi
\providecommand{\bibinfo}[2]{#2}
\providecommand{\eprint}[2][]{\url{#2}}

\bibitem{xu2018splitting}
\bibinfo{author}{Xu, Y.} \& \bibinfo{author}{Goodacre, R.}
\newblock \bibinfo{journal}{\bibinfo{title}{{On Splitting Training and
  Validation Set: A Comparative Study of Cross-Validation, Bootstrap and
  Systematic Sampling for Estimating the Generalization Performance of
  Supervised Learning}}}.
\newblock {\emph{\JournalTitle{Journal of analysis and testing}}}
  \textbf{\bibinfo{volume}{2}}, \bibinfo{pages}{249--262}
  (\bibinfo{year}{2018}).

\bibitem{kuhn2013applied}
\bibinfo{author}{Kuhn, M.}, \bibinfo{author}{Johnson, K.} \emph{et~al.}
\newblock \emph{\bibinfo{title}{Applied predictive modeling}},
  vol.~\bibinfo{volume}{26} (\bibinfo{publisher}{Springer},
  \bibinfo{year}{2013}).

\bibitem{guyon1997scaling}
\bibinfo{author}{Guyon, I.} \emph{et~al.}
\newblock \bibinfo{journal}{\bibinfo{title}{A scaling law for the
  validation-set training-set size ratio}}.
\newblock {\emph{\JournalTitle{AT\&T Bell Laboratories}}}
  \textbf{\bibinfo{volume}{1}} (\bibinfo{year}{1997}).

\bibitem{refaeilzadeh2009cross}
\bibinfo{author}{Refaeilzadeh, P.}, \bibinfo{author}{Tang, L.} \&
  \bibinfo{author}{Liu, H.}
\newblock \bibinfo{journal}{\bibinfo{title}{Cross-validation.}}
\newblock {\emph{\JournalTitle{Encyclopedia of database systems}}}
  \textbf{\bibinfo{volume}{5}}, \bibinfo{pages}{532--538}
  (\bibinfo{year}{2009}).

\bibitem{litjens2017survey}
\bibinfo{author}{Litjens, G.} \emph{et~al.}
\newblock \bibinfo{journal}{\bibinfo{title}{A survey on deep learning in
  medical image analysis}}.
\newblock {\emph{\JournalTitle{Medical image analysis}}}
  \textbf{\bibinfo{volume}{42}}, \bibinfo{pages}{60--88}
  (\bibinfo{year}{2017}).

\bibitem{ker2017deep}
\bibinfo{author}{Ker, J.}, \bibinfo{author}{Wang, L.}, \bibinfo{author}{Rao,
  J.} \& \bibinfo{author}{Lim, T.}
\newblock \bibinfo{journal}{\bibinfo{title}{{Deep Learning Applications in
  Medical Image Analysis}}}.
\newblock {\emph{\JournalTitle{IEEE Access}}} \textbf{\bibinfo{volume}{6}},
  \bibinfo{pages}{9375--9389} (\bibinfo{year}{2017}).

\bibitem{anwar2018medical}
\bibinfo{author}{Anwar, S.~M.} \emph{et~al.}
\newblock \bibinfo{journal}{\bibinfo{title}{{Medical Image Analysis using
  Convolutional Neural Networks: A Review}}}.
\newblock {\emph{\JournalTitle{Journal of medical systems}}}
  \textbf{\bibinfo{volume}{42}}, \bibinfo{pages}{1--13} (\bibinfo{year}{2018}).

\bibitem{maier2018rankings}
\bibinfo{author}{Maier-Hein, L.} \emph{et~al.}
\newblock \bibinfo{journal}{\bibinfo{title}{Why rankings of biomedical image
  analysis competitions should be interpreted with care}}.
\newblock {\emph{\JournalTitle{Nature communications}}}
  \textbf{\bibinfo{volume}{9}}, \bibinfo{pages}{1--13} (\bibinfo{year}{2018}).

\bibitem{wang2019deep}
\bibinfo{author}{Wang, H.}, \bibinfo{author}{Won, D.} \& \bibinfo{author}{Yoon,
  S.~W.}
\newblock \bibinfo{journal}{\bibinfo{title}{A deep separable neural network for
  human tissue identification in three-dimensional optical coherence tomography
  images}}.
\newblock {\emph{\JournalTitle{IISE Transactions on Healthcare Systems
  Engineering}}} \textbf{\bibinfo{volume}{9}}, \bibinfo{pages}{250--271}
  (\bibinfo{year}{2019}).

\bibitem{butola2020deep}
\bibinfo{author}{Butola, A.} \emph{et~al.}
\newblock \bibinfo{journal}{\bibinfo{title}{Deep learning architecture
  “{LightOCT}” for diagnostic decision support using optical coherence
  tomography images of biological samples}}.
\newblock {\emph{\JournalTitle{Biomedical Optics Express}}}
  \textbf{\bibinfo{volume}{11}}, \bibinfo{pages}{5017--5031}
  (\bibinfo{year}{2020}).

\bibitem{irmak2021multi}
\bibinfo{author}{Irmak, E.}
\newblock \bibinfo{journal}{\bibinfo{title}{Multi-classification of brain tumor
  {MRI} images using deep convolutional neural network with fully optimized
  framework}}.
\newblock {\emph{\JournalTitle{Iranian Journal of Science and Technology,
  Transactions of Electrical Engineering}}} \textbf{\bibinfo{volume}{45}},
  \bibinfo{pages}{1015--1036} (\bibinfo{year}{2021}).

\bibitem{sadad2021brain}
\bibinfo{author}{Sadad, T.} \emph{et~al.}
\newblock \bibinfo{journal}{\bibinfo{title}{Brain tumor detection and
  multi-classification using advanced deep learning techniques}}.
\newblock {\emph{\JournalTitle{Microscopy Research and Technique}}}
  \textbf{\bibinfo{volume}{84}}, \bibinfo{pages}{1296--1308}
  (\bibinfo{year}{2021}).

\bibitem{yagis2021effect}
\bibinfo{author}{Yagis, E.} \emph{et~al.}
\newblock \bibinfo{journal}{\bibinfo{title}{Effect of data leakage in brain
  {MRI} classification using {2D} convolutional neural networks}}.
\newblock {\emph{\JournalTitle{Scientific reports}}}
  \textbf{\bibinfo{volume}{11}}, \bibinfo{pages}{1--13} (\bibinfo{year}{2021}).

\bibitem{butola2019volumetric}
\bibinfo{author}{Butola, A.} \emph{et~al.}
\newblock \bibinfo{journal}{\bibinfo{title}{Volumetric analysis of breast
  cancer tissues using machine learning and swept-source optical coherence
  tomography}}.
\newblock {\emph{\JournalTitle{Applied optics}}} \textbf{\bibinfo{volume}{58}},
  \bibinfo{pages}{A135--A141} (\bibinfo{year}{2019}).

\bibitem{kermany2018large}
\bibinfo{author}{Kermany, D.}, \bibinfo{author}{Zhang, K.} \&
  \bibinfo{author}{Goldbaum, M.}
\newblock \bibinfo{journal}{\bibinfo{title}{{Large Dataset of Labeled Optical
  Coherence tomography (OCT) and Chest X-Ray images}}}.
\newblock {\emph{\JournalTitle{Mendeley Data}}} \textbf{\bibinfo{volume}{3}},
  \bibinfo{pages}{10--17632} (\bibinfo{year}{2018}).

\bibitem{srinivasan2014fully}
\bibinfo{author}{Srinivasan, P.~P.} \emph{et~al.}
\newblock \bibinfo{journal}{\bibinfo{title}{Fully automated detection of
  diabetic macular edema and dry age-related macular degeneration from optical
  coherence tomography images}}.
\newblock {\emph{\JournalTitle{Biomedical optics express}}}
  \textbf{\bibinfo{volume}{5}}, \bibinfo{pages}{3568--3577}
  (\bibinfo{year}{2014}).

\bibitem{Retinalkaggle}
\bibinfo{title}{{Retinal OCT Images} (optical coherence tomography)}.
\newblock
  \bibinfo{howpublished}{\url{https://kaggle.com/paultimothymooney/kermany2018}}.
\newblock \bibinfo{note}{Accessed: 2022-02-10}.

\bibitem{micko2021diagnosis}
\bibinfo{author}{Micko, A.} \emph{et~al.}
\newblock \bibinfo{journal}{\bibinfo{title}{Diagnosis of pituitary adenoma
  biopsies by ultrahigh resolution optical coherence tomography using neuronal
  networks}}.
\newblock {\emph{\JournalTitle{Frontiers in Endocrinology}}}
  \bibinfo{pages}{1345} (\bibinfo{year}{2021}).

\bibitem{najeeb2018classification}
\bibinfo{author}{Najeeb, S.} \emph{et~al.}
\newblock \bibinfo{title}{Classification of retinal diseases from {OCT} scans
  using convolutional neural networks}.
\newblock In \emph{\bibinfo{booktitle}{2018 10th International Conference on
  Electrical and Computer Engineering (ICECE)}}, \bibinfo{pages}{465--468}
  (\bibinfo{organization}{IEEE}, \bibinfo{year}{2018}).

\bibitem{chen2021classification}
\bibinfo{author}{Chen, Y.-M.}, \bibinfo{author}{Huang, W.-T.},
  \bibinfo{author}{Ho, W.-H.} \& \bibinfo{author}{Tsai, J.-T.}
\newblock \bibinfo{journal}{\bibinfo{title}{Classification of age-related
  macular degeneration using convolutional-neural-network-based transfer
  learning}}.
\newblock {\emph{\JournalTitle{BMC bioinformatics}}}
  \textbf{\bibinfo{volume}{22}}, \bibinfo{pages}{1--16} (\bibinfo{year}{2021}).

\bibitem{latha2021automated}
\bibinfo{author}{Latha, V.}, \bibinfo{author}{Ashok, L.} \&
  \bibinfo{author}{Sreeni, K.}
\newblock \bibinfo{title}{{Automated Macular Disease Detection using Retinal
  Optical Coherence Tomography images by Fusion of Deep Learning Networks}}.
\newblock In \emph{\bibinfo{booktitle}{2021 National Conference on
  Communications (NCC)}}, \bibinfo{pages}{1--6} (\bibinfo{organization}{IEEE},
  \bibinfo{year}{2021}).

\bibitem{kermany2018identifying}
\bibinfo{author}{Kermany, D.~S.} \emph{et~al.}
\newblock \bibinfo{journal}{\bibinfo{title}{Identifying medical diagnoses and
  treatable diseases by image-based deep learning}}.
\newblock {\emph{\JournalTitle{Cell}}} \textbf{\bibinfo{volume}{172}},
  \bibinfo{pages}{1122--1131} (\bibinfo{year}{2018}).

\bibitem{tsuji2020classification}
\bibinfo{author}{Tsuji, T.} \emph{et~al.}
\newblock \bibinfo{journal}{\bibinfo{title}{Classification of optical coherence
  tomography images using a capsule network}}.
\newblock {\emph{\JournalTitle{BMC ophthalmology}}}
  \textbf{\bibinfo{volume}{20}}, \bibinfo{pages}{1--9} (\bibinfo{year}{2020}).

\bibitem{kamran2019optic}
\bibinfo{author}{Kamran, S.~A.}, \bibinfo{author}{Saha, S.},
  \bibinfo{author}{Sabbir, A.~S.} \& \bibinfo{author}{Tavakkoli, A.}
\newblock \bibinfo{title}{{Optic-Net: A Novel Convolutional Neural Network for
  Diagnosis of Retinal Diseases from Optical Tomography Images}}.
\newblock In \emph{\bibinfo{booktitle}{2019 18th IEEE International Conference
  On Machine Learning And Applications (ICMLA)}}, \bibinfo{pages}{964--971}
  (\bibinfo{organization}{IEEE}, \bibinfo{year}{2019}).

\bibitem{athanasiou2019deep}
\bibinfo{author}{Athanasiou, L.~S.}, \bibinfo{author}{Olender, M.~L.},
  \bibinfo{author}{Jos{\'e}, M.}, \bibinfo{author}{Ben-Assa, E.} \&
  \bibinfo{author}{Edelman, E.~R.}
\newblock \bibinfo{title}{A deep learning approach to classify atherosclerosis
  using intracoronary optical coherence tomography}.
\newblock In \emph{\bibinfo{booktitle}{Medical Imaging 2019: Computer-Aided
  Diagnosis}}, vol. \bibinfo{volume}{10950}, \bibinfo{pages}{163--170}
  (\bibinfo{organization}{SPIE}, \bibinfo{year}{2019}).

\bibitem{wang2021deep}
\bibinfo{author}{Wang, C.} \emph{et~al.}
\newblock \bibinfo{journal}{\bibinfo{title}{Deep-learning-aided forward optical
  coherence tomography endoscope for percutaneous nephrostomy guidance}}.
\newblock {\emph{\JournalTitle{Biomedical optics express}}}
  \textbf{\bibinfo{volume}{12}}, \bibinfo{pages}{2404--2418}
  (\bibinfo{year}{2021}).

\bibitem{gesperger2020improved}
\bibinfo{author}{Gesperger, J.} \emph{et~al.}
\newblock \bibinfo{journal}{\bibinfo{title}{Improved diagnostic imaging of
  brain tumors by multimodal microscopy and deep learning}}.
\newblock {\emph{\JournalTitle{Cancers}}} \textbf{\bibinfo{volume}{12}},
  \bibinfo{pages}{1806} (\bibinfo{year}{2020}).

\bibitem{saratxaga2021characterization}
\bibinfo{author}{Saratxaga, C.~L.} \emph{et~al.}
\newblock \bibinfo{journal}{\bibinfo{title}{{Characterization of Optical
  Coherence Tomography Images for Colon Lesion Differentiation under Deep
  Learning}}}.
\newblock {\emph{\JournalTitle{Applied Sciences}}}
  \textbf{\bibinfo{volume}{11}}, \bibinfo{pages}{3119} (\bibinfo{year}{2021}).

\bibitem{singla2019automated}
\bibinfo{author}{Singla, N.}, \bibinfo{author}{Dubey, K.} \&
  \bibinfo{author}{Srivastava, V.}
\newblock \bibinfo{journal}{\bibinfo{title}{Automated assessment of breast
  cancer margin in optical coherence tomography images via pretrained
  convolutional neural network}}.
\newblock {\emph{\JournalTitle{Journal of biophotonics}}}
  \textbf{\bibinfo{volume}{12}}, \bibinfo{pages}{e201800255}
  (\bibinfo{year}{2019}).

\bibitem{chetoui2020deep}
\bibinfo{author}{Chetoui, M.} \& \bibinfo{author}{Akhloufi, M.~A.}
\newblock \bibinfo{title}{Deep retinal diseases detection and explainability
  using {OCT} images}.
\newblock In \emph{\bibinfo{booktitle}{International Conference on Image
  Analysis and Recognition}}, \bibinfo{pages}{358--366}
  (\bibinfo{organization}{Springer}, \bibinfo{year}{2020}).

\bibitem{thomas2021novel}
\bibinfo{author}{Thomas, A.} \emph{et~al.}
\newblock \bibinfo{journal}{\bibinfo{title}{A novel multiscale and multipath
  convolutional neural network based age-related macular degeneration detection
  using {OCT} images}}.
\newblock {\emph{\JournalTitle{Computer Methods and Programs in Biomedicine}}}
  \textbf{\bibinfo{volume}{209}}, \bibinfo{pages}{106294}
  (\bibinfo{year}{2021}).

\bibitem{rasti2017macular}
\bibinfo{author}{Rasti, R.}, \bibinfo{author}{Rabbani, H.},
  \bibinfo{author}{Mehridehnavi, A.} \& \bibinfo{author}{Hajizadeh, F.}
\newblock \bibinfo{journal}{\bibinfo{title}{Macular {OCT} classification using
  a multi-scale convolutional neural network ensemble}}.
\newblock {\emph{\JournalTitle{IEEE transactions on medical imaging}}}
  \textbf{\bibinfo{volume}{37}}, \bibinfo{pages}{1024--1034}
  (\bibinfo{year}{2017}).

\bibitem{farsiu2014quantitative}
\bibinfo{author}{Farsiu, S.} \emph{et~al.}
\newblock \bibinfo{journal}{\bibinfo{title}{Quantitative classification of eyes
  with and without intermediate age-related macular degeneration using optical
  coherence tomography}}.
\newblock {\emph{\JournalTitle{Ophthalmology}}} \textbf{\bibinfo{volume}{121}},
  \bibinfo{pages}{162--172} (\bibinfo{year}{2014}).

\bibitem{karimian2018deep}
\bibinfo{author}{Karimian, N.}, \bibinfo{author}{Salehi, H.~S.},
  \bibinfo{author}{Mahdian, M.}, \bibinfo{author}{Alnajjar, H.} \&
  \bibinfo{author}{Tadinada, A.}
\newblock \bibinfo{title}{Deep learning classifier with optical coherence
  tomography images for early dental caries detection}.
\newblock In \emph{\bibinfo{booktitle}{Lasers in Dentistry XXIV}}, vol.
  \bibinfo{volume}{10473}, \bibinfo{pages}{1047304}
  (\bibinfo{organization}{International Society for Optics and Photonics},
  \bibinfo{year}{2018}).

\bibitem{wang2020oct}
\bibinfo{author}{Wang, R.} \emph{et~al.}
\newblock \bibinfo{title}{{OCT} image quality evaluation based on deep and
  shallow features fusion network}.
\newblock In \emph{\bibinfo{booktitle}{2020 IEEE 17th International Symposium
  on Biomedical Imaging (ISBI)}}, \bibinfo{pages}{1561--1564}
  (\bibinfo{organization}{IEEE}, \bibinfo{year}{2020}).

\bibitem{bussola2021ai}
\bibinfo{author}{Bussola, N.}, \bibinfo{author}{Marcolini, A.},
  \bibinfo{author}{Maggio, V.}, \bibinfo{author}{Jurman, G.} \&
  \bibinfo{author}{Furlanello, C.}
\newblock \bibinfo{title}{{AI} slipping on tiles: {Data leakage in digital
  pathology}}.
\newblock In \emph{\bibinfo{booktitle}{International Conference on Pattern
  Recognition}}, \bibinfo{pages}{167--182} (\bibinfo{organization}{Springer},
  \bibinfo{year}{2021}).

\bibitem{khan2021global}
\bibinfo{author}{Khan, S.~M.} \emph{et~al.}
\newblock \bibinfo{journal}{\bibinfo{title}{A global review of publicly
  available datasets for ophthalmological imaging: barriers to access,
  usability, and generalisability}}.
\newblock {\emph{\JournalTitle{The Lancet Digital Health}}}
  \textbf{\bibinfo{volume}{3}}, \bibinfo{pages}{e51--e66}
  (\bibinfo{year}{2021}).

\bibitem{maqc2010microarray}
\bibinfo{author}{Consortium, M.} \emph{et~al.}
\newblock \bibinfo{journal}{\bibinfo{title}{The microarray quality control
  (maqc)-ii study of common practices for the development and validation of
  microarray-based predictive models}}.
\newblock {\emph{\JournalTitle{Nature biotechnology}}}
  \textbf{\bibinfo{volume}{28}}, \bibinfo{pages}{827} (\bibinfo{year}{2010}).

\bibitem{jurman2012comparison}
\bibinfo{author}{Jurman, G.}, \bibinfo{author}{Riccadonna, S.} \&
  \bibinfo{author}{Furlanello, C.}
\newblock \bibinfo{journal}{\bibinfo{title}{A comparison of mcc and cen error
  measures in multi-class prediction}}.
\newblock {\emph{\JournalTitle{PLOS ON}}}  (\bibinfo{year}{2012}).

\bibitem{chicco2020advantages}
\bibinfo{author}{Chicco, D.} \& \bibinfo{author}{Jurman, G.}
\newblock \bibinfo{journal}{\bibinfo{title}{The advantages of the matthews
  correlation coefficient (mcc) over f1 score and accuracy in binary
  classification evaluation}}.
\newblock {\emph{\JournalTitle{BMC genomics}}} \textbf{\bibinfo{volume}{21}},
  \bibinfo{pages}{1--13} (\bibinfo{year}{2020}).

\bibitem{chicco2021matthews}
\bibinfo{author}{Chicco, D.}, \bibinfo{author}{T{\"o}tsch, N.} \&
  \bibinfo{author}{Jurman, G.}
\newblock \bibinfo{journal}{\bibinfo{title}{The matthews correlation
  coefficient (mcc) is more reliable than balanced accuracy, bookmaker
  informedness, and markedness in two-class confusion matrix evaluation}}.
\newblock {\emph{\JournalTitle{BioData mining}}} \textbf{\bibinfo{volume}{14}},
  \bibinfo{pages}{1--22} (\bibinfo{year}{2021}).

\bibitem{sokolova2009systematic}
\bibinfo{author}{Sokolova, M.} \& \bibinfo{author}{Lapalme, G.}
\newblock \bibinfo{journal}{\bibinfo{title}{A systematic analysis of
  performance measures for classification tasks}}.
\newblock {\emph{\JournalTitle{Information processing \& management}}}
  \textbf{\bibinfo{volume}{45}}, \bibinfo{pages}{427--437}
  (\bibinfo{year}{2009}).

\end{thebibliography}

\end{document}